\documentclass[12pt]{article}
\def\d{\partial}
\def\nn{\nonumber\\}

\def\cd{\cdot}

\def\a{\alpha}

\def\h1{\hat 1}
\def\h3{\hat 3}
\def\h2{\hat 2}

\def\ln{\,{\rm ln\,}}

\def\g{\gamma}

\def\G{\Gamma}

\def\ln{\,{\rm ln\,}}
\def\O{\Omega}

\def\S{\Sigma}
\def\s{\sigma}

\def\na{\nabla}

\def\f{\frac}
\def\s{\sigma}

\def\h0{\hat 0}

\def\q{\quad}
\def\qq{\qquad}
\def\nn{\nonumber\\}

\def\bl{\biggl}

\def\br{\biggr}

\begin{document}

\begin{center}
    {\large\bf Space-time structure and fields of bound charges}\nn

\bigskip

    {\bf S.A. Podosenov}\\
 {\it All-Russian Scientific-Research Institute for Optical and Physical
   Measurements, Ozernaya 46, Moscow 119361, Russia}

    e-mail: podosenov@mail.ru
\end{center}
\bigskip

\begin{abstract}
   An exact solution for the field of a charge in a uniformly accelerated
   noninertial frame of reference (NFR) alongside the "Equivalent Situation
   Postulate" allows one to find space-time structure as well as fields from
   arbitrarily shaped charged conductors, without using Einstein's equations.
   In particular, the space-time metric over a charged plane can be related
   to the metric being obtained from an exact solution to Einstein-Maxwell's
   equations. This solution describes an equilibrium of charged dust in
   parallel electric and gravitational fields.
The field and metric outside a conducting ball have been found. The method
proposed eliminates divergence of the proper energy and makes classical
electrodynamics consistent at any sufficiently small distances. An experiment
is proposed to verify the approach suggested.
   \end{abstract}

\begin{center}
   {\bf 1. Introduction}
\end{center}

   The space-time curvature in a physical theory is conventionally related
   to solutions of Einstein's equations. However, the transition to a rigid
   NFR also result in a bend of the spacetime, as shown in [1] and
   repeated in [2], [3]. In the present paper a non-Euclidean form of space-
   time is shown to be included by electric fields as well.

   Let us formulate the

   {\bf Equivalent Situation Postulate}

  {\it The field of a point charge, being at equilibrium in a constant
  electric field, is equivalent to the field from this charge in a uniformly
   accelerated NFR, if the constraint reaction forces, accelerating the charge
   and holding it immovable, are equal}.

   This postulate is absent in classical electrodynamics where the field of
   a point charge being at rest in an inertial frame of reference (IFR) is
   Coulomb's spherically symmetric one, irrespective of whether the charge
   is free or the sum of forces acting on it is zero. The field from a
   uniformly accelerated charge proves to be axially symmetric for the
   observer in an NFR.

   Thus, the identical physical situation, in which the charges are, results
   in the field with different symmetry! We try to solve this paradox in
   the present paper.

\begin{center}
  {\bf 2. Electrostatic Field of Bound Charges}
\end{center}

   In paper [1] a rigid uniformly accelerated motion is shown to be
   realizable in the Riemannian space of a constant curature with the metric
\begin{eqnarray}
dS^2=\exp(2{a_0}y^1/c^2)(dy^0)^2-(dy^1)^2-(dy^2)^2-(dy^3)^2. 
\end{eqnarray}
   where the acceleration $a_0$ is directed along the $y^1$ axis. One
   independent component of the curvature tensor calculated by metric (1)
   is of the form
\begin{eqnarray}
R_{10,10}=-\frac{{a_0}^2}{c^4}\exp(2{a_0}{y^1}/c^2).   
\end{eqnarray}

   We use the "Equivalent Situation Postulate" and mathematical formalism of
   NFR to find the electric field of charged conductors. Charges on the
   surface of a conductor experience negative pressure forces due to the
   external field. The metal lattice induces a force preventing the charges
   from escaping the body surface. Thus, the field of the system under
   consideration is equivalent to the field of the system of uniformly
   accelerated charges whose accelerations are directed to the inside of
   the conductor. To find the total field of a charged conductor, one should
   sum the fields from charges of surface elements.

   The equation for a scalar potential $A_0$ of a point charge located
   at the origin of coordinates (1) has the form
$$
\triangle{A_0}-\frac{a_o}{c^2}\frac{{\d}{A_0}}{{\d}{y^1}}=
-4\pi{Q}\exp\biggl(\frac{a_0y^1}{c^2}\biggr)\delta(y^1)\delta(y^2)
\delta(y^3).       \eqno(3)
$$

   This equation can be obtained from Maxwell's equations written in the
   generally covariant form [4] similar to the case of a "given gravitational
   field". The solution to the equation results in the relation
$$
A_0=\frac{Q}{r}\exp\biggl\{-\frac{{a_0}{r}(1-\cos\theta)}
{2c^2}\biggr\}.   \eqno(4)
$$

   For the electric field strength $\vec E$ we have
$$
\vec E=\frac{Q}{r^2}\exp\biggl\{-\frac{{a_0}{r}(1-\cos\theta)}
{2c^2}\biggr\}\biggl[\frac{\vec r}{r}+\frac{{a_0}{r}}{2c^2}
\biggl(\frac{\vec r}{r}-\vec i\biggr)\biggr],  \eqno(5)
$$
   where $r$ is a three-dimensional (Euclidean) distance from the origin of
   coordinates, coinciding with the charge, to the point of observation;
   $\theta $ is an angle between the radius vector $\vec r$ and $\vec i$,
$\vec i=\vec a_0/{\mid{\vec a_o}\mid}$.

\begin{center}
{\bf3. Field of a Charged Plate}
\end{center}

Calculate the field being created by an infinite charged metallic plate of
the whose thickness $h\to 0$. On both sides of the plate the charge
density will
be constant and equal to $\sigma$.
   A contribution to the scalar
   potential $A_0$ from the whole plate can be calculated by integrating the
   contribution from elementary charges $dQ$ of each of the plate side in a
   flat space (but not in a Riemannian space-time).

We match the origin of coordinates to the
plate centre by directing the $y^1$ axis, perpendicular to the plate, towards
its upper side. After integration we find the potential in the upper half-
space, with the plate charge density assumed equal to $\sigma_0=2\sigma$.
$$
A_0=-\frac{2\pi\sigma_0{c^2}}{{a_0}}
\biggl[1-\exp\biggl\{-\frac{{a_0}{y^1}}
{c^2}\biggr\}\biggr].       \eqno(6)
$$

The value of the field intensity of the charged plane is
$$
E_1=F_{01}=-\frac{{\d}{A_0}}{{\d}{y^1}}=2\pi\sigma_0
\exp\biggl\{-\frac{{a_0}{y^1}}
{c^2}\biggr\}.       \eqno(7)
$$
Define the space-time metric by choosing a charged dust over the plane of
like charges as the basis of a noninertial frame of reference (NFR). Let each
of the particles be connected with the charged plane by a weightless and
inextensible filament. The charged dust in this self-similar problem sets
the basis of the frame of reference (FR) whose structure is determined by a
mutual solution to the equations of "motion" and Maxwell's equations. The
solutions of the latter is given by formula (7). Dust particles will be fixed
so that the deformation velocity tensor and the rotation velocity tensor be
equal to zero, and the filament tensions be nonzero. Let the plane be
infinite, the charge density on it be constant, and the charge-to-mass ratio
be identical for all particle of the basis. By definition, we do not consider
the dust to create a field, but the field is determined only by charges of the
plane according to (1). It is evident that the physical situation for dust
particles in the upper half-space is equivalent to the situation in an NFR
"moving" downwards in the direction of the plane, which is described
mathematically by metric (1) with a negative sign in the exponent, i. e. by
substituting $a\to a_0$. The values $a_0$ and $a$ are to be calculated.

The interval element will be sought in a more detailed form than (1)
$$
dS^2=\exp(\nu(y^1)){dy^0}^2-\exp(\lambda(y^1)){dy^1}^2-{dy^2}^2-{dy^3}^2,
                       \eqno(8)
$$
where the functions $\nu(y^1)$ and $\lambda(y^1)$ are to be found.
To find them, we take
advantage of the solution to Maxwell's three-dimensional equations whose
form is not dissimilar to Maxwell's equations in a given gravitational field.
The solution to the equation for the induction vector $D^1$ outside the plane
has the form $D^1=const\exp(-\lambda/2)$. Using both the relations
arising from [4] and formula (7),
we find
$$
\frac{\nu+\lambda}{2}=-\frac{a_0y^1}{c^2}.  \eqno(9)
$$
The second equation connecting the functions $\nu$ and $\lambda$ will be found
from the equations of "motion"
$$
F^1=-\frac{q}{m_0c^2}F^{10}V_0, \eqno(10)
$$
where $V_0=\exp(\nu/2)$ is the zero component of the 4-velocity
of a basis particle of the
mass $m_0$, charge $q$ in the Lagrangian comoving NFR. The minus sign in (10)
means that the first curvature vector of a basis particle $F^1$ is determined
only by the filament tension acting in the direction opposite to that of the
force due to the field. On the other hand, $F^1$ may be found
directly from the
form of metric (8) as the only nonzero component of the 4-acceleration in the
Lagrangian comoving NFR. This results in the relation.
$$
\frac{1}{2}\frac{{\d}\nu}{{\d}y^1}=-\frac{E_0q}{m_0c^2}\exp(\lambda/2),
\eqno(11)
$$
where $E_0=2\pi\sigma_0$. The solution to system (9), (10), providing
$y^1=0$, $\nu=0$, has the
form
$$
\exp(\nu/2)=1-\frac{E_0q}{m_0a_0}\biggl(1-\exp\biggl(-\frac{a_0y^1}{c^2}
\biggr)\biggr),
\qquad \exp(\lambda/2)=\exp\biggl(-\frac{\nu}{2}-\frac{a_0y^1}{c^2}\biggr).
\eqno(12)
$$

Metric (8) admits an evident simplification, if, instead of $y^1$, one
introduces another Lagrangian coordinate by the formula
$$
\tilde y^1=\int_0^{y^1}\exp(\lambda/2)\,dy^1
$$

As a results, we obtain
$$
dS^2=\exp\biggl(-\frac{2E_0q\tilde y^1}{m_0c^2}\biggr){dy^0}^2
-{d\tilde {y}^{1}}^2-{dy^2}^2-{dy^3}^2.
\eqno(13)
$$

This formula is similar to (1) with a negative acceleration $a=-E_0q/m$
directed
towards the plane, if the plane charge and the test one $q$ are of like
signs, and with a positive acceleration for unlike charges. The space-time
will be flat, if a test particle is neutral. Thus, contrary to General
Relatity (GR), the metric depends not only on a field strength of the charge
creating the field, but also on the value and sign of a test charge in the
field.

If for particles of the basis the test-charge-to-mass ratio is identical
for all particles, and the charges of the plane and test particle are unlike,
then metric (13) is identical to (1) with $a=a_0$.

   To compare theory with experiment, we express the fields in terms of
   "physical" and tetrad components. Choose the comoving tetrads for the
   metric so that the vector $\vec e_{(0)}$ should be directed along
   the time line,
   and the triad $\vec e_{(k)}$ - along the coordinate axes $y^k$
   (Lames gauge
   [5], [6]), where the tetrad indices enclosed in brackets.
$$
e_{(\alpha)}^\mu=\frac{\delta_\alpha^\mu}{\sqrt{\vert g_{\alpha\alpha}
\vert}}, \qquad e^{(\alpha)}_\mu=\delta_\mu^\alpha
\sqrt{\vert g_{\alpha\alpha}\vert},
\qquad e^{(0)}_\mu=V_\mu, \qquad e^\mu_{(0)}=V^\mu,
\eqno(14)
$$
   without summing over $\alpha$. For the "physical" components
   of the field tensor we find
$$
F_{(0)(1)}=E_{(1)}=e^\mu_{(0)}e^\nu_{(1)}F_{\mu\nu}=2\pi\sigma_0
=const. \eqno(15)
$$

   From (15) it follows that the field of the charged plane in terms of local
   tetrads is the same as that conventionally considered in Cartesian
   coordinates. However, the space-time geometry being due to an
   electrostatic field proved to be pseudo-Riemannian.

   We calculate the field created by an electron "suspended" by a filament
   over
   a positively charged plane. The situation with the electron of a mass
   $m$ and a charge $e$ "suspended" over the plane is equivalent to its
   putting into NFR (1) moving with the acceleration directed along
   the axis $y^1$ and equal to $a_0=eE/m$. The calculation by formula (5) of
   the field intensity of the plane $E=100$ $kV/m$ for
   field-aligned points differs
   from field-transverse ones by 10\% at the distance 1 m from the charge.

   While performing a standard calculation, the field of the electron should
   be isotropic and Coulomb's, like the field of a free electron.

   It can be shown that the correction is not small for negatively charged
   bodies as well. A force $F_e=eE/2$ acts on each of the electrons
   in the field.
   The force induced by the metal lattice is oppositely directed, and equal
   to $F_e$, in value. This force induces the "acceleration" $a_0=eE/(2m)$.
   For conductors charged positively, on the surface there are not electrons
   but positive ions with the masses much more than those of electrons.
   Therefore its "acceleration" $a_0$, and hence the field correction will
   be much less than for electrons.

   Propose an experiment to verify the approach suggested.

   Set a problem of calculating the capacity of a flat capacitor without
   regard for boundary effects.

   The capacitor capacity is calculated by the formula
$$
C={Q^2}/{2W},                \eqno(15a)
$$
   where $Q$ is a charge, and $W$ is a total energy between the plates.
   The energy density $\rho $ of the electromagnetic field corresponds to
   the $T_{(0)(0)}=T^{(0)(0)}=T^{(0)}_{(0)}$ components of the
   energy momentum tensor of the
   electromagnetic field in curvilinear coordinates has the form [4]
$$
T_{\mu\nu}=\frac{1}{4\pi}\biggl(-F_{\mu\beta}F_{\nu}{}^\beta+
\frac{1}{4}F_{\beta\gamma}F^{\beta\gamma}g_{\mu\nu}\biggr),
\eqno(16)
$$
   where the tetrad components of the tensor are calculated by the rule
$$
T_{(\mu)(\nu)}=e^\alpha_{(\mu)}e^\beta_{(\nu)}T_{\alpha\beta}.
\eqno(17)
$$

The field energy may be calculated according [6] by the formula
$$
W=\int\sqrt{-g}T^{\mu\nu}V_{\nu}\,dS_\mu, \eqno(18)
$$
   where $g$ is the metric tensor determinant, $dS_\mu=V_{\mu}dV$. $dS_\mu$
   is a
   geometrical object equal to the product of an element of the
   hypersurface, orthogonal to the basis world lines and based on three
   infinitesimally small displacements, by a unit vector  of the normal
   ( i. e. the 4-velocity $V_\mu$).
Because relation (18) is of
importance, we discuss it in more detail.

Since we consider the electromagnetic field energy-momentum tensor (16)
outside charges creating the field, the relation should be valid
$$
\na_{\mu}T^{\mu\nu}=0.
\eqno(19)  
$$

In a flat space-time in Galilean coordinates this relation is the
electromagnetic field energy-momentum conservation law. In Riemannian
spacetime this relation, generally, is not a conservation law at all, since a
covariant derivative of the energy-momentum tensor enters in it instead of a
partial one, which is well-known from the literature on GR.

While calculating the energy in (18), we, in fact, use not the "conservation
law" (19), but the real conservation law resembling the charge conservation
law.

This follows from the equality
$$
\na_{\mu}\biggl(T^{\mu\nu}V_{\nu}\biggr)=V_{\nu}\na_{\mu}T^{\mu\nu}+
T^{\mu\nu}\na_{\mu}V_{\nu}=T^{\mu\nu}V_{\mu}F_{\nu}=0.
\eqno(20)
$$

In deriving (20) we have allowered for the congruence of world lines of FR
basis particles, determined by the equation of "motion" (10), being
irrotational and rigid, the vectors $V_\mu$ and $F_\mu$ being orthogonal,
and the
convolution of the antisymmetric field tensor $F_{\mu\nu}$ with
the product of
first curvature vectors being equal to zero.

Relation (20) may be rewritten in the form
$$
\na_{\mu}T^{\mu(0)}=\f{1}{\sqrt{-g}}\f{\d}{\d y^{\mu}}\biggl(
\sqrt{-g}T^{\mu(0)}\biggr)=0.
\eqno(21)
$$

We integrate expression (21) over an invariant 4-volume
$$
\int\sqrt{-g}\na_{\mu}T^{\mu(0)}d^4y=\int\f{\d}{\d y^{\mu}}\biggl(
\sqrt{-g}T^{\mu(0)}\biggr)d^4y=0.
\eqno(22)
$$

Using the Gauss theorem and assuming that on the "lateral" timelike
hypersurface, enclosing a spatial volume, the integral vanishes (which is
valid in problems with charges located in a finite volume), we obtain
$$
\oint\sqrt{-g}T^{\mu(0)}dS_{\mu}=\int_{V_1}\sqrt{-g}T^{(0)(0)}d^3y
-\int_{V_2}\sqrt{-g}T^{(0)(0)}d^3y=0.
\eqno(23)
$$

Whence follows the conservation law of the "charge"- type quantity
$$
\int_{V_1}\sqrt{-g}T^{(0)(0)}d^3y
=\int_{V_2}\sqrt{-g}T^{(0)(0)}d^3y,
\eqno(24)
$$
wherein $V_1$ and $V_2$ are the three-dimensional
volumes occupied by the field
at different instants.

The comparison of (18) with (24) implies an identity of the quantities that
represent the electromagnetic field energy.

Calculate the field energy of a negatively charged plane. In the standard
calculation the plane field energy of the field strength $E$, confined within
a cylinder with the base area  $S$ and height $h$ on each side of the plane
with the elements of the cylinder perpendicular to the plane, is evidently
equal to
$$
W_0=\f{E^2}{8\pi}2hS=\sigma_0Qh,\qq Q=\sigma_0S.
\eqno(25)
$$
For $h\to\infty$, $W_0\to\infty$.

In this case the field energy in the above-mentioned volume
$$
W=\int\sqrt{-g}T^{(0)(0)}\,dV=2\frac{E^2{S}c^2}{8\pi{a_0}}
\biggl(1-\exp\biggl(-\frac{a_0{h}}{c^2}\biggr)\biggr),\eqno(26)
$$
where $E=2\pi\sigma_0$.

It is evident that at small distances $h$ from the plane, i. e. for
$$
\frac{a_0{h}}{c^2}\ll{1}
$$
the field energy in  volume calculated according to (26) coincides with the
classical expression (25). However, although in the classical consideration
for $h\to\infty$, $W_0\to\infty$, in this case the field energy
inside an infinitely long
cylinder remains a finite quantity defined by the equality
$$
W=\int\sqrt{-g}T^{(0)(0)}\,dV=2\frac{E^2{S}c^2}{8\pi{a_0}}
=\frac{E{S}c^2m}{2\pi{e}}=\f{Qmc^2}{e}=Nmc^2. \eqno(27)
$$
Thus:

{\sl The energy of the electric field inside a long cylinder proved
to be equal
to the rest energy of $N$ electrons located on a charged surface of the area
$S$ inside the cylinder. The charge $Q$ of an area element does not enter in
this energy}.

Formula (27) remains valid for a positively charged plane as well. In this
case the role of a mass (since positrons  are not stable) plays the mass of
an atom on the conductor that lost one electron.

{\sl Thus, the proper energy of charges on the plane proved to be equal
to their
rest energy!}

   The field energy between the capacitor plates can be calculated
    by the formula
$$
W=\int\sqrt{-g}T^{(0)(0)}\,dV=\frac{E^2{S}c^2}{8\pi{a_0}}
\biggl(1-\exp\biggl(-\frac{a_0{d}}{c^2}\biggr)\biggr)
\eqno(28)
$$
   where $S$ is a plate area, $d$ is a distance between the plates.

   Assuming $u=Ed$, where $u$ is a potential difference between the
   plates, we find the capacitor capacity $C$ in the form
$$
C={Q^2}/{2W}=\frac{S}{4\pi{d}}\frac{eu}{2mc^2}\frac{1}{1-\exp
\biggl(-\frac{eu}{2mc^2}\biggr)}.   \eqno(29)
$$

   In the approach suggested the capacitor capacity increases with
   increasing the voltage applied to the capacitor. For example, at the
   voltage $5\cdot10^4$ $V$ the capacity increase should amount to  2.4\%.

   \begin{center}
{\bf 4. Geometry of Uniformly Accelerated NFR\\ and Einstein-Maxwell's
   Equations}
\end{center}

   Although metric (13) is not immediately related to solving
   Einstein's equations, the presence of curvature stimulates to seek
   relations to General Relativity. Let us relate metric (13) to an exact
   solution of Einstein-Maxwell's equations. For the Einstein tensor
   components

$$G_{\mu\nu}=R_{\mu\nu}-\frac{1}{2}g_{\mu\nu}R $$

we find
\setcounter{equation}{29}
\begin{eqnarray}
G_{00}=0,\quad G_{11}=0,\quad G_{0k}=0, \quad G_{kl}=0,\nn
\quad (k\ne{l})
\quad (k,l=1,2,3) \quad
G_{22}=G_{33}=\frac{1}{2}R=\frac{{a_0}^2}{c^4}.  
\end{eqnarray}

   Write down Einstein's equations allowing for "cosmological constant
   $\Lambda$" we obtain
\begin{eqnarray}
G_{\mu\nu}+\Lambda g_{\mu\nu}= \kappa T_{\mu\nu}, \qquad
\kappa=\frac{8\pi k}{c^4} 
\end{eqnarray}
   where $k$ is the gravitational constant.
   Choose the electromagnetic field energy-momentum tensor (16) as a source.
   Due to the equality $T^\nu_\nu=0$ for the "cosmological constant"
   we obtain
\begin{eqnarray}
\Lambda=\frac{{a}^2}{2c^4}=\frac{{E_0}^2q^2}{2m^2c^4}
=\frac{(2\pi\sigma_0)^2q^2}{2m^2c^4}=const.  
\end{eqnarray}

A consistent solution to Einstein-Maxwell's set results in a relation
   between masses and charges of test particles of the basis
\begin{eqnarray}
\frac{q^2}{m^2}=2k. 
\end{eqnarray}

   The metric obtained can be simply interpreted. The basis of such an
   NFR is the charged dust being at equilibrium in parallel homogeneous
   electric and gravitational fields. (The dust itself therewith does not
   create fields). The masses of dust particles are $\sqrt{2}$ times less
   than those of stable elementary black holes, "maximons" [8]. The particle
   with a proton charge has a mass close to $10^{-6}$ g which $10^{21}$
   more than the electron mass.
\begin{center}
{\bf 5. Centrally Symmetric Electrostatic Field}
\end{center}

Consider an electrostatic field having a central symmetry. This field may
be created by a charged spherical body or  a point charge.
 The electrostatic field of bound charges differs from that of free
charges. Consider a charged conducting ball of radius $R$ and charge $Q$.
It is necessary to find the electric field strength $\vec E$ and
determine the
space-time geometry outside the ball.

Each of the charges of the conductor will be on the ball surface and will
experience a "negative pressure" force being due to the external field, which
they create, directed along the external normal to the surface [7]. This
force is compensated by the force induced by the lattice holding charges on
the sphere surface. Thus, the physical situation under consideration is
equivalent to that in which there are charges connected with weightless
filaments of length $R$ fixed at a common centre. Consequently, the field
created by each of the charges will be such as if each of the charges would
be uniformly accelerated and its acceleration would be directed to the ball
centre.

In accordance with (4)  the potential $dA_0$ at the point of
observation has the form
\begin{eqnarray}
dA_0=\frac{dQ}{r'}\exp\biggl\{-\frac{{a_0}{r'}(1-\cos\theta)}
{2c^2}\biggr\}.   
\end{eqnarray}
where $r'$ is a three-dimensional (Euclidean) distance from the charge
$dQ$ to the point of observation, $\theta$ is an angle between the
radius vector $\vec {r}'$ and $\vec i$, $\vec i=\vec a_0/\mid\vec a_0\mid$.
$\vec i$ is directed to the sphere
centre for an element $dQ$ of the charge. The value of the "acceleration"
$a_0$ for charges of the negatively charged sphere (electrons) is calculated
by the formula
\begin{eqnarray}
a_0=\frac{eE}{2m}.    
\end{eqnarray}
 Here $e$ is the charge value,
$m$ is the electron mass, $E$ is a field strength on the sphere surface.
For the ball charged positively the "relativistic" effect will be much less
pronounced than for the negatively charged one, because the mass $m$ of
 the
positive ion  exceeding
the electron mass. Hence the field from a positively charged ball almost
coincides with the classical one, and for a negatively charged ball the
"relativistic" corrections may prove to be considerable.

Integrating (34), we obtain
\begin{eqnarray}
A_0=\frac{Q\exp(-\zeta^2)\sqrt{\pi}i}{4R\zeta}
\biggl[\Phi(i\zeta(1-2R/r))-\Phi(i\zeta)\biggr],
\qquad \zeta=\frac{r}{R}\sqrt{\frac{eQ}{8mc^2R}}.
\end{eqnarray}

In (36) $r$ is a distance from the ball centre to the point of observation,
$\Phi(\zeta)$ is the probability integral
$$
\Phi(i\zeta)=\frac{2i}{\sqrt{\pi}}\int_0^\zeta\exp(t^2)\,dt.
$$
The set of electrons on the sphere surface does not belong to the congruence
of world lines of the NFR basis (formula (1) ) but are included
in the set of world lines belonging to the Lagrangian comoving spherically
symmetric NFR with the metric
\begin{eqnarray}
dS^2  = \exp(\nu)(dy^0)^2- r^2(d\theta^2+\sin^{2}\theta{d}\phi^2)
- \exp(\lambda)(dr)^2.
\end{eqnarray}

The functions $\nu(r)$ and $\lambda(r)$ need to be determined. To find them,
we resort to a solution to Maxwell's spherically symmetric static equations,
using metric (37), similar to the notation of the equations of electrodynamics
in a "given gravitational field" [4]. Then compare the solution obtained
with the expression for the field derived from (36). As a result, we find the
equation of a constraint on the functions $\nu(r)$ and $\lambda(r)$
\begin{eqnarray}
\exp\biggl(\frac{\nu+\lambda}{2}\biggr)
=\frac{\exp(-\zeta^2)\sqrt{\pi}}{4\delta}\biggl\{\biggl[\Phi(i\zeta-
2i\delta)
-\Phi(i\zeta)\biggr](1+2\zeta^2)i\nn
-\frac{2\zeta\exp(\zeta^2)}{\sqrt{\pi}}
\bigl[1-\exp(-4\delta(\zeta-\delta))\bigr]\biggr\},
\qquad \delta=\sqrt{\frac{eQ}{8mc^2R}}. 
\end{eqnarray}

To find the second equation connecting these functions, we consider a force,
being due to the field, acting on a test charge $q$ fixed at the point
$r$ from the ball centre. Let the mass of the test charge be $m_0$. Then
the first curvature vector $F^1$ of the world line of this charge can be
found from the relation
$$
F^1=\frac{1}{2}\frac{d\nu}{dr}\exp(-\lambda)
$$
and  from the force acting on the charge due to the constraint holding it
immovable in the field. This force is numerically equal to the force due
to the field and is opposite to it in sign.
\begin{eqnarray}
F^1=\frac{1}{2}\frac{d\nu}{dr}\exp(-\lambda)=
-\frac{q}{m_0c^2}F^{10}V_0=-\frac{Qq}{m_0c^2r^2}\exp(-\lambda/2).
\end{eqnarray}

As a result, we find
\begin{eqnarray}
\exp(\nu/2)=-\frac{q}{m_0c^2}\int {F_{01}}\,dr=-\frac{Q
q\sqrt{\pi}}{4Rm_0c^2}\int\biggl\{\exp(-\zeta^2)\biggl[\Phi(i\zeta-
2i\delta)
-\Phi(i\zeta)\biggr]\nn(1/\zeta^2+2)i
-\frac{2}{\zeta\sqrt{\pi}}
\bigl[1-\exp(-4\delta(\zeta-\delta))\bigr]\biggr\}\,d\zeta+▒_1.
\end{eqnarray}

We determine the integration constant $C_1$ from the requirement of space
being Euclidean at infinity.
Performing the integration in (40) analytically presents no problems, since
by definition
$$
F_{01}=-\frac{{\d}A_0}{{\d}r}
$$
where $A_0$ is determined from (36). As a result, we obtain
\begin{eqnarray}
\exp(\nu/2)=1+\frac{Qq\exp(-\zeta^2)\sqrt{\pi}i}{m_0c^24r\delta}
\biggl[\Phi(i\zeta-2i\delta)
-\Phi(i\zeta)\biggr]
=1+\frac{qA_0}{m_0c^2}.  
\end{eqnarray}

From (38) and (41) we have
\begin{eqnarray*}
\exp\biggl(\frac{\lambda}{2}\biggr)=\frac{\frac{\exp(-\zeta^2)\sqrt{\pi}}
{4\delta}\biggl\{\biggl[\Phi(i\zeta-
2i\delta))
-\Phi(i\zeta)\biggr](1+2\zeta^2)i\biggr\}}
{1+\frac{Qq\exp(-\zeta^2)\sqrt{\pi}i}{m_0c^24r\delta}
\biggl[\Phi(i\zeta-2i\delta)-\Phi(i\zeta)\biggr]}
\end{eqnarray*}
$$
-\frac{\frac{\zeta}{2\delta}
\bigl[1-\exp(-4\delta(\zeta-\delta))\bigr]}
{1+\frac{Qq\exp(-\zeta^2)\sqrt{\pi}i}{m_0c^24r\delta}
\biggl[\Phi(i\zeta-2i\delta)-\Phi(i\zeta)\biggr]}=
-\frac{\frac{r^2}{Q}\frac{{\d}A_0}{{\d}r}}
{1+\frac{qA_0}{m_0c^2}}. 
 \eqno(42)
$$

Analyze (41), (42) expanding the expressions as a series in a dimensionless
parameter $\delta$. It should be mentioned that for a negatively charged
metallic ball $\phi=50$ $kV$ corresponds to $\delta=0.11$ for an electron with the
classical radius $e^2/mc^2$, $\delta=0.35355$ for a proton $\delta=0.015$.

For a metallic ball charged up to $\phi=50 kV$ the value of
$\delta\sim 10^{-4}$  (a concrete
value depends on the atomic weight of the metal).

The expansion in terms of $\delta $ with retaining terms proportional to
the first power of $\delta$ (in the numerator of (42) we expand into a
series retaining terms with $\delta^2$) leads to the relation
\setcounter{equation}{42}
\begin{eqnarray}
\exp(\nu)=\biggl(1+\frac{Qq}{rm_0c^2}\biggr)^2=\exp(-\lambda).
\end{eqnarray}
In formula (43) the charges $Q$ and $q$ can be of equal as well as
opposite signs. As distinct from GR, the space-time metric depends both
on the charge $Q$ creating a field and on the value of a test charge $q$
. If as the test charges one chooses equal charges whose Coulomb's
repulsion force is exactly equal to Newton's attraction force, then such
particles will be classically noninteracting. If, besides, one equates
Newton's gravitational force between charges on the sphere to its Coulomb's
electrostatic repulsion, then evident relations $q^2/m_0^2=k$, $Q^2/M^2=k$
will be
valid, where $M$ is a mass of charges $Q$, $k$ is the gravitational
constant. (The solution is presented in SGSE units). Metric (43) in
this case coincides with an exact electrovacuum spherically symmetric
consistent solution to Einstein's and Maxwell's equations and is called
Reissner-Nordstr\"om's metric. Estimate the value of $\delta $ at which an
approximate solution (i.e. Reissner-Nordstr\"om's)  differs least from
the exact one (41), (42).

Choose the electron charge $e$ as the least charge. Then from the equality
$e^2/M^2=k$ we find $M=1.86\cdot10^{-6}$ $g$.
Considering the particle to be spherical and
assuming its density to be equal to that of nuclear matter $\rho=2.7\cdot
10^{14}$
$g/cm^{3}$, we find $\delta=2.5\cdot10^{-9}$ in accordance with (24). Since we have
obtained Reissner-Nordstr\"om's solution as an expansion in terms of $ \delta
$, it almost coincides with the exact one (41), (42) for the last example.
It should be noted that the nonzero tetrad component $F_{(0)(1)}$ of the field
tensor exactly corresponds to Coulomb's nonconventional expression for the
field outside a charged ball in Minkowski space for any $\nu $ and $\lambda$.

If the metallic ball is positively charged and test charges of the FR basis
are negative, then metric (43) for such a system, to a high accuracy, is
similar to Schwarzschild's metric in GR. It follows from a formal
substitution $Qq\to -kMm_0$ and retaining two terms of the expansion,
while being
squared. As distinct to Schwarzschild's metric, $g_{00}>0$
for any distance from the
ball centre.

Propose an experiment to verify the approach suggested.

Calculate the capacitance $C$ of a charged metallic ball of radius $R$.
When
considering classically, we use the formula $C=R$. In our case the
capacitance
will depend on the potential applied to the ball as well as on the charge
sign. Calculate the capacitance by the formula
\begin{eqnarray}
C={Q^2}/{2W}.
\end{eqnarray}

The energy density $\rho$ of the electromagnetic field corresponds to the
$T_{(0)(0)}=T^{(0)(0)}=T^{(0)}_{(0)}$ components of
the energy-momentum tensor in terms of comoving tetrads
 (14) or
\begin{eqnarray}
\rho=\frac{Q^2}{8\pi{r^4}},
\end{eqnarray}
which coincides with the expression for the field energy density outside the
charged sphere in Minkowski space.

However, to find the total fields energy outside the charged ball, it is
necessary to integrate over space in the Riemannian space-time with metric
(37) determined from (41), (42). For that we use formula (18)
yielding
\begin{eqnarray}
W=\int\sqrt{-g}T^{\mu\nu}V_{\nu}\,dS_\mu=\frac{Q^2}{2}\int_R
^\infty\frac{\exp\bigl(\frac{\lambda+\nu}{2}\bigr)}{r^2}\,dr=\frac
{A_0(R)Q}{2}, 
\end{eqnarray}
where $A_0(R)$ is determined from (36) ar $r=R$.

As a result, we obtain for the capacitance $C$
\begin{eqnarray}
▒=R\biggl[\frac{2i\delta}{\sqrt{\pi}\exp(-\delta^2)\Phi(i\delta)
}\biggr], 
\end{eqnarray}

For example, at the potential equal to $- 50$ $kV$ an addition to the capacitance
should amount to 0.8\%.

To calculate the ball capacitance, one may, instead of formulae (30), (33),
to determine the capacitance conventionally by the formula
\begin{eqnarray}
C=\frac{Q}{A_0},
\end{eqnarray}
where $A_0$ is calculated from (22) at $r=R$.
Thus, both formulae (44) and (48) proved to be equivalent as well as in the
classical consideration.

As known [4], in classical relativistic mechanics finite dimensions cannot
be assigned to elementary particles, and they should be considered to be
pointlike. The charged particle under such consideration has an infinite
proper energy as well as a mass. The physical senselessness of this result
calls for the fundamental principles of electrodynamics itself to be bounded
by certain limits.

We prove that in the model proposed the above-mentioned difficulty is absent,
and the proper energy of a negative point charge $Q$ containing $N$
electrons corresponds to the quantity $W=2Nmc^2$.

To prove it, we use expressions (36) and (46) for $R$. Assuming  $Q=Ne$
and making use of the well-known asymptotic formula [9]
\begin{eqnarray}
\sqrt{\pi}z\exp(z^2)(1-\Phi(z))\sim 1+\sum_{m=1}^\infty(-1)^m
\frac{1\cdot{3}\ldots(2m-1)}{(2z^2)^m},
\end{eqnarray}
we find for $z=\delta\gg{1}$ the expression
$$
A_0(R)\approx \frac{Q}{2R\delta^2}\biggl(1+\frac{1}{2\delta^2}\biggr)=
\frac{4mc^2}{e}\biggl(1+\frac{4mc^2R}{eQ}\biggr)
$$
$$
W=\frac{QA_0}{2}=2Nmc^2\biggl(1+\frac{4mc^2R}{e^2N}\biggr).
$$

Whence for $R\to {0}$ we have
\begin{eqnarray}
W=2Nmc^2.
\end{eqnarray}

The quantity of energy of a point particle proved to be independent of the
sign and value of the charge. From (50) it follows that the field energy of
a negatively charged particle with the charge $Q=Ne$ is determined
by the total
rest energy of its $2N$ electron masses.

In particular, for one electron "smeared" over a sphere of
radius the $R\to 0$, the
proper field energy coincides with its rest energy $W=2mc^2$.
The energy of a point particle does not
depend on the sign and value of the charge.

In the "nonrelativistic" approximation $\delta\to 0$ we
have classical relations
$$
A_0(r)=\frac{Q}{r},\qquad W=\frac{Q^2}{2R}
$$
for the potential and energy of a charged conducting ball respectively.

From formula (50) it follows
that the mass of the charged particle is completely electromagnetic. Thus,
the approach we suggest widens the area of application of the classical field
theory as one goes to sufficiently small distances.

\begin{center}
{\bf 6. Axially Symmetric Field}
\end{center}

Consider an axially symmetric field being created by a thin infinitely long
charged metallic cylinder.

We match the $z$ axis to the cylinder one, having chosen the origin of
coordinates at the cylinder centre. The calculation of the field will be
performed in a cylindrical system of coordinates, using formula (46) to find
the potential, where $r'$ is the three-dimensional (Euclidean) distance from
the charge $dQ$ to the observation point; $\theta$ is the angle between the
radius vector $\vec r$ and $\vec i$, $\vec i=
\vec {a_0}/{\mid{\vec {a_0}}\mid}$. $\vec i$ for each element of
the charge $dQ$ is
directed towards the centre of the cylinder perpendicular to its surface. The
quantity of "acceleration" $a_0$ for charges of a negatively
charged cylindrical
surface (electrons) is calculated by formula (35).

Each of the charges on the conductor will be located on the cylinder surface
and experiences a "negative pressure" force (due to the field) created by
them, in the direction of an external normal to the surface [7]. This force
is compensated by the force induced by the lattice holding charges on the
cylinder surface. Thus, the physical situation under consideration is
equivalent to that in which the charges are connected by weightless filaments
of length $R$ equal to the cylinder radius fixed along the cylinder axis.
Hence the field created by each of the charges will be the same as if each
charge would move with an acceleration directed towards the cylinder axis.
According to (1), each of the electrons on the cylinder belongs to a tangent
flat space but Riemannian space-time. Hence the operation of integration
over the cylinder occurs in a flat space and is correct.

From geometrical considerations it is easy to obtain a formula relating the
quantities $'r$ and $\theta$ to the radial  $\rho$ and angular  $\phi$
cylindrical coordinates.
$$
r'\cos\theta=R-\rho\cos\phi.
\eqno(51)
$$

The charge element $dQ$ on the cylinder surface is representable in the form
$$
dQ=\s R d\phi dz=\f{\g}{2\pi}d\phi dz,
\eqno(52)
$$
where $\sigma$ and $\gamma$ are the surface and linear densities of charges.

To simplify calculations, we turn from a thin cylinder to a charged filament
assuming $R\to 0$ and considering $\gamma$ to be a constant
finite quantity. As a
result, for the potential $A_0$ we obtain the expression
$$
A_0=\f{\g}{\pi}\int_0^{\pi} d\phi\exp\bl(-\f{\rho a_0\cos\phi}{2c^2}\br)
\int_{-\infty}^{+\infty}\f{\exp\bl(-\f{a_0 \sqrt{\rho^2+z^2}}{2c^2}\br)}
{\sqrt{\rho^2+z^2}}\,dz.
\eqno(53)
$$

The calculation of the integrals leads to the relation
$$
A_0=2\g I_0(\a)K_0(\a),\qq \a=\f{\rho a_0}{2c^2}=\f{\rho e E_0}{4mc^2}.
\eqno(54)
$$

In the latter relation $E_0$ is a field strength on the cylinder surface,
$I_0$,
$K_0$ are modified Bessel functions in the conventional designations [9].

We find space-time geometry outside the charged filament and calculate the
field energy created by filament charges.

The set of charges on the cylinder surface does not belong to the congruence
of NFR basis world lines (1) but is included in the set of basis world lines
belonging to the Lagrangian cylindrically symmetric comoving NFR with the
metric of the form
$$
dS^2  =\exp(\nu)(dy^0)^2- \exp(\lambda)d\rho^2-\rho^2 d\phi^2-dz^2,
\eqno(55)
$$
where $\nu$ and $\lambda$ depend only on $\rho$.

The functions $\nu(\rho)$ and $\lambda(\rho)$ are to be found.
To find them, we take
advantage of the solution to Maxwell's cylindrically symmetric static
equations using metric (55). The form of these equations similar to that of
the equations of electrodynamics in a "given gravitational field" [4]. Then
we shall compare the solution obtained with the expression for the field
being obtained from (54).

For a nonzero radial component of the "induction" $D^1$ we have
$$
\frac{1}{\sqrt\gamma}\frac{\d}{{\d}\rho}\biggl(\rho\exp
(\lambda/2)D^1\biggr)=0,           
\eqno(56)
$$
whose solution will be
$$
D^1=\frac{2\g}{\rho}\exp(-\lambda/2).           
\eqno(57)
$$
In (56) and (57) between the "induction" $D^1$, the field intensity $E^1$
and the
field tensor component $F_{01}$ there exists the well-known relations
$$
D_1=\gamma_{11}D^1=\frac{2\g}{\rho}\exp(\lambda/2)=\frac{1}
{\sqrt{g_{00}}}E_1,\q E_1=F_{01},\q \gamma_{kl}=-g_{kl} ,
\eqno(58)
$$
where $\gamma_{kl}$ is the spatial metric tensor with the
determinant equal to
$\gamma$.

From (54) we find the nonzero components $F_{01}=-F_{10}$
$$
F_{01}=-\frac{\d A_0}{\d\rho}=\f{\g a_0}{c^2}(K_1(\a)I_0(\a)-I_1(\a)K_0(\a)).
\eqno(59)
$$

Equating (59) to the expression of $F_{01}$ from (58),
we find the equation that
connects the functions $\nu(\rho)$ and $\lambda(\rho)$
$$
\exp\biggl(\frac{\nu+\lambda}{2}\biggr)=\a(K_1(\a)I_0(\a)-I_1(\a)K_0(\a))=
\f{\rho}{2\g}F_{01}.
\eqno(60)
$$

To find the second equation connecting these functions, we consider a force
(due to the field) acting on the test charge $q$ fixed at the point with the
coordinate $\rho$ reckoned from the filament axis. Let the test charge mass
be $m_0$. Then the first curvature vector $F^1$ of this charge
world line may be
found from relation
$$
F^1=\frac{1}{2}\frac{d\nu}{d\rho}\exp(-\lambda). 
\eqno(61)
$$

This quantity may also be found from the force acting on the charge due to
the constraint holding the charge fixed in the field. This force is
numerically equal to the force due to the field and is opposite to it in
sign.
$$
F^1=\frac{1}{2}\frac{d\nu}{d\rho}\exp(-\lambda)=
-\frac{q}{m_0c^2}F^{10}V_0=
-\frac{2\g q}{m_0c^2\rho}\exp(-\lambda/2). 
\eqno(62)
$$

From (59)-(62) we find
$$
\exp(\nu/2)=-\frac{q}{m_0c^2}\int {F_{01}}\,d\rho=\f{qA_0}{m_0c^2}+▒_1,
\eqno(63)
$$
where $C_1$ is an integration constant.

The integration constant will be found by considering the asymptotic $A_0$ at
long distances from the filament. It is well-known that, while considering
classically, the potential $A_0$ diverges logarithmically at long distances
from the filament.

In the case under consideration it is not valid. Prove this.

In tetrads (14), based on formulae (57), (58), the electromagnetic field
tensor component $F_{(0)(1)}$ has the form
$$
F_{(0)(1)}=\frac{2\g}{\rho}.           
\eqno(64)
$$

On the cylinder surface the quantity $E_0$ coincides with the tetrad
component  $F_{(0)(1)}$. Hence for the quantity $\alpha$ we have
$$
 \a=\f{\rho a_0}{2c^2}=\f{\rho e E_0}{4mc^2}=\f{\rho e \g}{R2mc^2},
 \q \rho\ge R.
\eqno(65)
$$

From the analysis of (65) it follows that $\alpha\to \infty$
for $\rho\to\infty$. At long distances
from the filament one may use the well-known expansion of Bessel functions
[9]
$$
 I_0(\a)\simeq\sqrt{\f{1}{2\pi\a}}e^{\a},\q K_0(\a)\simeq\sqrt
 {\f{\pi}{2\a}}e^{-\a}.
\eqno(66)
$$

For the 4-component of the potential $A_0$ this results in the expression
$$
 A_0(\a)=\f{\g}{\a}=\f{2mc^2R}{e\rho}.
\eqno(67)
$$
Whence it follows that for $\rho\to\infty$, $A_0\to 0$.

Thus, we have proved that in this case the behaviour of $A_0$ at infinity
differs sharply from the classical analogue: instead of a diverging quantity
we have obtained a vanishing quantity.

This provides a possibility of determining the integration constant $C_1$ from
the requirement of a space being Euclidean at infinity. As a result, from
(63) we have
$$
\exp(\nu/2)=1+\f{qA_0}{m_0c^2}.
\eqno(68)
$$

Using (60), we find
$$
\exp(\lambda/2)=-\f{\rho}{2\g}\f{\d A_0}{\d\rho}\f{1}{1+\f{qA_0}{m_0c^2}}.
\eqno(69)
$$

For weak fields, with $\alpha\ll 1$, using the well-known expansions
for modified
Bessel functions [9]
$$
 I_0(\a)\simeq 1,\q K_0(\a)\simeq -(\ln(\a/2)+C),
\eqno(70)
$$
we find
$$
 A_0(\a)\simeq -2\g\bl(\ln\bl(\f{\rho a_0}{2c^2}\br)+C\br),
 \q -\f{\d A_0}{\d\rho}\simeq\f{2\g}{\rho},
\eqno(71)
$$
where $C=0.577...$ is the Euler constant.

Expression (71), within an accuracy of the value of a constant in the
potential, coincides with the classical expression. However, the space-time
metric even in the case of a weak field remains Riemannian, is determined by
formula (68), and instead of (69) we have
$$
\exp(\lambda/2)=\f{1}{1+\f{qA_0}{m_0c^2}}.
\eqno(72)
$$

Calculate the field energy of an element of the charged filament of
length $h$. For the tetrad component $T^{(0)(0)}$
of the energy-momentum tensor we
have an expression
$$
T^{(0)(0)}=\f{\g^2}{2\pi\rho^2},
\eqno(73)
$$
\begin{eqnarray*}
W=\int\sqrt{-g}T^{\mu\nu}V_{\nu}\,dS_\mu=
\f{h\g}{2}\int_R^\infty F_{01}\,d\rho
\end{eqnarray*}
$$
=-\f{h\g}{2}\int_R^\infty \f{\d A_0}{\d\rho}\,d\rho=
\frac{A_0(R)\g h}{2}-\frac{A_0(\infty)\g h}{2}.
\eqno(74)
$$

Since, according to (67) $A_0(\infty)=0$
and the filament, by the data given, has
a radius $R\to 0$, relation (74), based on (65), has the form
$$
W=\frac{A_0(R)\g h}{2}=\g^2 hI_0\bl(\frac{e\g}{2mc^2}\br)
K_0\bl(\frac{e\g}{2mc^2}\br).
\eqno(75)
$$

From formula (75) it follows that the field energy, inside the cylinder of
length $h$ and infinite radius, is finite, which differs sharply from the
classical value of charged filament energy that is infinitely high.

For close packing of charges on the filament, i. e. providing that
$$
\frac{e\g}{2mc^2}\gg 1
\eqno(76)
$$
we have an expression for the energy
$$
W=\frac{\g hmc^2}{e}=Nmc^2,
\eqno(77)
$$
where $N$ is the number of electrons along the filament length $h$.

{\sl Thus, for three kinds of symmetry (flat, spherical or cylindrical)
the field
energy being created by point charges does diverge, as in the classical case,
but is determined by the rest energy of charges creating the field. Therewith
the quantity of charge drops out of the energy}.

Thus, the approach we have proposed widens the field of applicability of the
classical field theory in proceeding to sufficiently small distances.

\begin{center}
{\bf 7. Integral Equation for a Density of Bound Charges}
\end{center}

To obtain an integral equation for the density $\sigma$ of bound charges
located on the surface of a conducting body of an arbitrary shape (the
surface assumed sufficently smooth for the differentiability conditions to be
satisfied), we divide the surface into area elements, the potential of each of
them at the observation point is given by the formula that follows from (34)
$$
dA_0=\frac{\s'dS'}{r'}\exp\biggl\{-\frac{{a_0'}{r'}(1-\cos\theta)}
{2c^2}\biggr\},   
\eqno(78)
$$
where $r'$ is the three-dimensional (Euclidean) distance of the charge
$dQ'=\s'dS'$ to
the observation point; $\sigma'$ is a current value of the charge density
depending on a point on the surface; $\theta$ is the angle between the radius
vector $\vec r$ and $\vec i$, $\vec i=\vec {a_0'}/{\mid{\vec {a_0'}}\mid}$
. $\vec i$ for each element of the charge $dQ'$ is directed
normally to the surface, towards the body interior. The quantity of
"acceleration" $a_0'$ for charges of a negatively charged body (electrons) is
calculated by the formula
$$
a_0'=\frac{eE'}{2m}=\f{2\pi\s'}{m}.                        
\eqno(79)
$$

Introduce, at each point of the surface, a unit vector $\vec n'$ in
the direction
of an external normal to the surface. Then relation (78) takes the form
$$
dA_0=
\frac{\s'dS'}{r'}\exp\biggl\{-\frac{\pi{e}\s'(r'+\vec n'\cd\vec r')}
{mc^2}\biggr\}.
\eqno(80)
$$

In the electrostatic case for any stationary metric there will be nonzero
$F_{0k}$ components of the electromagnetic field tensor. From (80),
integrating
over the surface and calculating the field tensor, at the observation point
with the coordinates $x^k$ we have an expression.
\footnote {Note that, although
each elementary charge on the body surface belongs to a curved space-time, a
spatial cross-section for each of the charges is Euclidean, since metric
(1) has a flat cross-section and at different (locally constant) values of
the acceleration. Hence, while calculating distances from integration area
elements for each of the charges to an arbitrary but fixed observation point,
the spatial metric tensor is the Kronecker symbol. For this reason, we do not
distinguish between covariant and contravariant spatial components.}
\begin{eqnarray*}
F_{0k}=-\f{\d A_0}{\d x^k}=\int\frac{\s'}{r'}
\exp\biggl\{-\frac{\pi{e}\s'(r'+\vec n'\cd\vec r')}{mc^2}\biggr\}\nn
\end{eqnarray*}
$$
\bl[\f{x^k-x'^k}{r'^2}+\f{\pi\s'e}{mc^2}\bl(\f{x^k-x'^k}{r'}+n'^k\br)\br]
\,dS'.
\eqno(81)
$$

Choose an observation point $x^k$ on the conductor surface and multiply the
expression obtained at the observation point by a unit vector of the normal
$\vec n$ to form the quantity being a scalar with respect to spatial
transforms $\Psi$.
\begin{eqnarray*}
\Psi=\int\frac{\s'}{r'}
\exp\biggl\{-\frac{\pi{e}\s'(r'+\vec n'\cd\vec r')}{mc^2}\biggr\}\nn
\end{eqnarray*}
$$
\bl[\f{\vec n\cd\vec N'}{r'}+\f{\pi\s'e}{mc^2}\bl(\vec n\cd\vec N'
+\vec n\cd\vec n'\br)\br]
\,dS'.
\eqno(82)
$$

Here $\vec N'$ is the unit vector along $\vec r'$ directed from
the integration element
to the observation point.

While conventionally considered, the quantity $\Psi$ is related to the charge
surface density $\sigma$ by the relation
$$
\Psi=2\pi\s.
$$

It is clear that in the "nonrelativistic" approximation we obtain Grinberg's
classical integral equation [10]. Taking account of the field of bound
charges leads to a more complicated dependence between $\Psi$ and $\sigma$.

Find restrictions on the space-time metric outside a charged conducting body
in such a way as we have done it for the charged ball.

Suspend  test charges (e. g., electrons) by weightless filaments in the field
created by this body. Since the forces due to the field are balanced by the
filament tension forces, then going into the Lagrangian comoving FR connected
with the charges, one may verify that the metric in such a system will be
equal to
$$
dS^2  =g_{00}(dy^0)^2- \g_{kl}dy^kdy^l.
\eqno(83)
$$
It is evident that the tensor components, due to a static nature of the field,
should not be dependent on the temporal coordinate $y^0$ but only on spatial
ones $y^k$. Since rotations are absent, $g_{0k}$ are equal to zero.
Since the suspended charges are fixed, the comovement conditions for metric
(83) should be satisfied as follows
$$
V^k=V_k=0,\q V^0=(g_{00})^{-1/2},\q V_0=(g_{00})^{1/2}.
\eqno(84)
$$
From the comovement conditions it follows that the first curvature vector
(4-acceleration) is not equal to zero, as in equilibrium in Minkowski space,
but is calculated by the well-known formula as in the case of the force
acting on a particle in the constant gravitational field [4]
$$
F^k=\G^k_{00}(V^0)^2=-\f{1}{2g_{00}}g^{km}\f{\d g_{00}}{\d y^m}.
\eqno(85)
$$
On the other hand, this quantity may also be found from the force acting on
a charge due to the constraint holding the charge to be fixed in the field.
This force is numerically equal to the force induced by the field and is
opposite to it in sign.
$$
F^k=-\f{1}{2g_{00}}g^{km}\f{\d g_{00}}{\d y^m}
=-\frac{e}{mc^2}F^{k0}V_0=\f{e}{mc^2}g^{km}F_{0m}V^0.
\eqno(86)
$$

The electric field from a charged body may also be found from Maxwell's
equations in a "given gravitational field" [4]
\footnote{It is natural that we
do not consider a true gravitational field at all, but simply use Maxwell's
equations in Riemannian space without using Einstein's equations.}
with an
unknown metric tensor for metric (83).
This equation outside the charged
body (the density of "suspended" test charges is neglected) has the form
$$
\f{\d}{\d y^k}\bl({\sqrt{\g}D^k}\br)=0, \q D^k=-\sqrt{g_{00}}F^{0k}.
\eqno(87)
$$
The field tensor (81) may be calculated if the charge density $\sigma$ on
the conductor surface is known. Since the conductor shape is known, one may
choose on it a two-dimensional orthogonal curvilinear coordinate frame with
coordinate spatial vectors in the direction of the normal to the body surface.
The fourth coordinate temporal vectors, according to (84), will be matched
to the direction of the 4-velocities of electrons being at rest on the
surface. On the basis of the given coordinate system, there arises a natural
tetrad system  with Lam{\'e}'s gauge. Since the field strength on the
conductor surface is normal to the surface, then the field tensor on it in
the selected system of coordinates has the only nonzero component, e. g.
$F_{01}$
$$
F_{01}=e^{(\mu)}_{0}e^{(\nu)}_{1}F_{(\mu)(\nu)}=\sqrt{g_{00}}\sqrt{\g_{11}}
F_{(0)(1)}.
\eqno(88)
$$

As for the above problems of a charged sphere and plane, we identify the
field tetrad component $F_{(0)(1)}$ with the charge density on the body.

The right side of relation (82) coincides in the new coordinate system
with the field tensor component $F_{01}$ on the conductor surface.
The field on
the conductor surface in a small neighbourhood of a point is formed from the
field being created by charges within this neighbourhood and by all other
charges. Knowing that the field on the surface from the charges outside the
neighbourhood is equal to half the whole field, we obtain an integral
equation to find the bound charge density in the form
\begin{eqnarray*}
2\pi\s\sqrt{g_{00}}\sqrt{\g_{11}}=\int\frac{\s'}{r'}
\exp\biggl\{-\frac{\pi{e}\s'(r'+\vec n'\cd\vec r')}{mc^2}\biggr\}\nn
\end{eqnarray*}
$$
\bl[\f{\vec n\cd\vec N'}{r'}+\f{\pi\s'e}{mc^2}\bl(\vec n\cd\vec N'
+\vec n\cd\vec n'\br)\br]
\,dS'.
\eqno(89)
$$
While deriving it, we have taken into account that the field about the above
point on the surface, created by all other charges on the body outside the
neighbourhood of the point, is continuous on the conductor boundary. The
field of the charges being within the neighbourhood has a break on the
conductor boundary. This field, combined with that of other charges outside
the conductor, near its surface, results in doubling the field of all other
charges outside the conductor and the total field vanishing inside the
conductor. Therefore, the total mean field on the conductor surface, equal to
the halfsum of the external and internal field, leads to the value
$2\pi\sigma$
instead of $4\pi\sigma$ in the left side of equality (89).

Equations (86), (87), (89)  are equations for finding the density
of a field and charges as well as the metric tensor components. To these
equations it is necessary to add the structure equations, obtained in paper
[1], connecting the Riemann-Christoffel tensor
$R_{\epsilon\sigma,\nu}^{\mu}$
 with the characteristics
of the contituum of charges "suspended" in the field
$$
R_{\epsilon\sigma,\nu}^{\mu}V_{\mu}=2\nabla_{[\epsilon}\Sigma_{\sigma]\nu}
+2\nabla_{[\epsilon}\Omega_{\sigma]\nu}
+2\nabla_{[\epsilon}(V_{\sigma]}F_\nu).            
\eqno(90)
$$
Here
$$
\Sigma_{\mu\nu}=\nabla_{(\mu}V_{\nu)}-V_{(\mu}F_{\nu)}, 
\eqno(91)
$$
$$
\Omega_{\mu\nu}=\nabla_{[\mu}V_{\nu]}-V_{[\mu}F_{\nu]},     
\eqno(92)
$$
$$
F_\mu=V^{\nu}\nabla_{\nu}V_\mu.     
\eqno(93)
$$

In the above formulae  $\Sigma_{\mu\nu}$ is the velocity deformation
tensor, $\Omega_{\mu\nu}$
is the rotation velocity tensor, $F_\mu$ are the first curvature vectors of
medium particle world lines.

The round brackets, enclosing the indices, serve as a symmetrization symbol,
and the squared brackets as an antisymmetrization one. The Greek indices
can take values from zero to three, the Latin ones from one to three.

From the form of metric (83) it follows that
$\S_{\mu\nu}=\O_{\mu\nu}=0$, and in the Lagrangian
comoving FR the structure equations are simplified reducing to the system
$$
R_{\epsilon\sigma,\nu 0}V^{0}=
2\nabla_{[\epsilon}(V_{\sigma]}F_\nu),            
\eqno(94)
$$
wherein the 4-velocity $V_{\s}$ and the 4-acceleration $F_{\nu}$ are
set according to
(84) and (85).

\begin{center}
{\bf 8. Quasinewtonian Centrally Symmetric Gravitational Field}
\end{center}

The Newtonian theory of gravitation is obtained from Einstein's General
Relativity (GR) equations providing all particle velocities are small and the
gravitational field is weak [4]. Einstein's equations in this case reduce to
the form
$$
R_0^0=\frac{\Delta\varphi}{c^2}=\frac{4\pi{k}\rho}{c^2},
$$
where $R_0^0$ is the temporal component of the Ricci tensor, $\Delta$ is the
Laplacian written in the same way as in a flat space, $\varphi$ is the
gravitational field potential, $c$ is the velocity of light in vacuo, $k$ is
the gravitational constant, $\rho$ is the mass density.

In the nonrelativistic limit $c\to\infty$ the written equation
coincides with
Poisson's one in Euclidean space with the unified time and the Riemann-
Christoffel curvature tensor, and hence the Ricci tensor, identically equal
to zero.

We should like to show that such a representation is not only possible and
that the Newtonian theory may be considered in Riemannian space-time.

The object of the paper is to establish a more intimate connection between
Newton's and Einstein's theories, considering the Newtonian theory together
with the structure equations that we have found.

In Minkowski space consider a centrally symmetric motion of a continuous
medium arising from the point at which the origin of coordinates is placed.
It is evident that for observers in the Lagrangian comoving FR the distance
between adjacent elements of the medium will vary with time, i. e. such a
frame is not rigid. In other words, a rigid radial motion of the contonuous
medium is impossible in Minkowski space.

In Riemannian space this situation is possible. It follows, e. g., from the
static equilibrium condition in a spherically symmetric gravitational field
described by Schwarzschild's metric [4]. For the observers being at rest on
the surface of an immovable gravitating sphere, from the GR viewpoint, the
acceleration is nonzero and directed away from the centre, perpendicular to
the surface, whereas the acceleration is equal to zero for the observers
holding the Newtonian viewpoint. And vice versa, a free falling body has a
nonzero acceleration in the Newtonian gravitational field and follows a
geodesic with the zero acceleration in Schwarzschild's field.

We seek a metric of the Lagrangial spherically symmetric comoving NFR in the
form (37), by analogy with GR.
NFR (37) is evidently rigid, since the metric
coeffficients are independent of time, and the component $g_{0k}$ being equal to
zero indicates an absence of rotations.

System
$$
 \nabla_{\mu}V_\nu=\Sigma_{\mu\nu}+\Omega_{\mu\nu}+V_{\mu}F_{\nu}
$$
 with regard to the formulated requirements as well as the
fulfilment of the comovement conditions
$$
V^k=V_k=0,\quad V^0=(g_{00})^{-1/2},\quad V_0=(g_{00})^{1/2},
\quad  F^1= F(r),\quad F^0=F^2=F^3=0
$$
reduces to one equation
$$
F^1=\frac{1}{2}\frac{d\nu}{dr}\exp(-\lambda).
\eqno(95)
$$

One may be sure that the structural equations (90) satisfy (95) without
additional constraints on the functions $\nu(r)$ and $\lambda(r)$.
Thus metric (37)
cannot be unambiguously determined from the given field of the first
curvature vectors $F^1$.

Consider some simplest possibilities. We perform a gedanken experiment.

{\bf a)} Let the observers being on the surface of the Earth,
whose rotation is not
taken into account, considering its density constant and assuming its shape
spherical, measure the gravitational field with accelerometers. They will
find that the acceleration fields is aligned with the radius, away from the
centre, perpendicular to the surface. To measure a field far from the
surface, we use a set of radial weightless bars along which the
accelerometers are set. The set of bars and accelerometers form a basis of
the uniformly accelerated rigid FR.

Farther and farther from the Earth's surface the acceleration field will
diminish following (in the zeroth approximation) the Newtonian law of
gravitation. If the observers consider their space to be flat and the law of
gravitation to be exact, then metric (37) will have the form [1]
$$
dS^2  = \exp(-r_g/r)(dy^0)^2- r^2(d\theta^2+\sin^{2}\theta{d}\phi^2)
- (dr)^2, 
\eqno(96)
$$
where $r_g=2kM/c^2$ is called a gravitational radius. While deriving
(96), we have
taken into account that $\lambda$, by definition of a flat space, and found
$\nu$ from (95) and the Newtonian law of gravitation.

Although the space metric is flat, the space-time metric (96) proved to be
Riemannian. Thus, the Newtonian theory of gravitation admits two logically
consistent interpretations in a flat space.

According to the convential interpretation, not only space but space-time is
flat in the Newtonian theory. Therewith two forces act on the body being on
the Earth's surface, viz. a gravitation force and reaction-at-support force,
which add up to zero and hence impart no acceleration  to the body.

In our interpretation, the only one force: the reaction-at-support force,
which imparts the body an acceleration being measured by the accelerometer
and calculated by formula (95) with the use of metric (96). If one removes
the support, the body will follow a geodesic in the space-time with metric
(96), whereas interpreted conventially in the absence of support the body
will move in a flat space-time under the action of the gravitational force.
The Newtonian treatment we have modified is closer to Einsteinian than to
pure Newtonian one. It can be shown, using [11], that the pericentre
displacement in one revolution, calculated in metric (96), is one-third as
large as that calculated in Schwarzschild's metric. The change of the
direction of light ray, while passing by the central body, calculated by (96)
is half as large as Schwarzschild's. Therefore, the model being proposed,
without claiming to substitute GR, establishes a more intimate connection
between the Newtonian and Einsteinian theories showing that the Newtonian
theory may be considered in Riemannian space-time. Although this
interpretation coincides with experimental data in the Newtonian
approximation, it does not allow for more fine effects, which are explained
by GR.

We try to modify the model in such a way that it should correspond more
precisely to the observational data.

{\bf b)} While deriving (96), it was assumed that $\lambda=0$, which
corresponds to
the flat spatial cross-section model. As a frame of reference outside the
Earth chosen was the system of rigid undeformable bars along which sound
propagates with infinitely large velocity to be in contradiction with a
finiteness of the interaction propagation velocity.
Therefore, to eliminate this disadvantage of the model, we shall consider as
in GR that the radially accelerated NFR basis structure outside the Earth is
equivalent to some elastic medium, being subject to strains, and hence
stresses, but having a zero deformation velocity tensor. From the form of
metric (37) it follows that we consider only small radial displacement of
the elastic medium for which only the radial component
$u_{rr}={\d}u_r/{\d}r$ of the deformation
tensor is nonzero (in designations of [12]) in spherical coordinates. As for
the components
$u_{\theta\theta}=u_{\phi\phi}=u_{r}/r$, they are negligible as compared
to $u_{rr}$, and are not taken
into account in the model under consideration.

It is convenient for the strain-stress relation to be determined in the
Lagrangian comoving NFR, considering the elastic medium to be shear-
stressless. For the NFR valid is Hooke's relativistic law [15] on the
hypersurface orhogonal to the world lines coinciding in form with the
classical expression [13]
$$
P^{ij}=\tilde \lambda{I_1}\gamma^{ij},\qquad I_1(\varepsilon)=
\gamma^{kl}\varepsilon_{kl}=\frac{1}{2}(1-\exp(-\lambda)), 
\eqno(97)
$$
where $I_1$ is the first invariant of the deformation tensor,
$\tilde \lambda$ is the
Lam{\'e} coefficient, $\gamma^{ij}=-g^{ij}$
is the spatial cross-section metric (37)
$$
\varepsilon_{kl}=\frac{1}{2}(\gamma_{ij}-\gamma'_{ij}),
$$
$\gamma'_{ij}$ is the metric tensor of a flat space in spherical coordinates.
The elastic medium should satisfy the continuity equation
$$
\nabla_{\mu}({\rho}V^{\mu})=0.
$$

The solution to the continuity equation results in the relation [13], [14]
$$
\rho=\rho_0\exp (-\lambda/2),
\eqno(98)
$$
where $\rho_0$ is the density of a "medium" in an undeformed state.

The equations of "motion" of the elastic medium in a Lagrangian NFR has the
form similar to its equilibrium conditions in the Newtonian gravitational
field [12] considered classically
$$
\nabla_j{P^{ij}}=-\rho_0{a^j},
\eqno(99)
$$
where $a^j$ are the "nonphysical" - affine components of the acceleration,
and
uplifting and lowering the tensor indices and calculating the covariant
derivative are performed with the aid of the spatial metric $\gamma_{ij}$.
Assuming
that the physical and tetrad components correspond to the Newtonian value,
from (98) and (99) in spherical coordinates we have an expression
$$
\exp(-\lambda)\frac{d\lambda}{dr}=-2\frac{\rho_0{kM}}
{\tilde \lambda{r}^2},
\eqno(100)
$$
whose
integration, providing that the space is flat ($\lambda=0$) at infinity,
results
in the relation
$$
\exp(-\lambda)=\biggl(1-\frac{2{kM}}{c_0^2{r}}\biggr),
\qquad c_0^2=\frac{\tilde \lambda}{\rho_0},
\eqno(101)
$$
where $c_0$ is the longitudinal sound velocity.

Taking into account that the first curvature vector
$F^1=А^{-2}a^1=(\gamma_{11})^{-1/2}kM/(cr)^2$, using (95) and
(101), we obtain an equation for $\nu$ whose integration, providing that
$\nu=0$
at infinity, gives
$$
\nu=2\biggl(\frac{c_0}{c}\biggr)^2\biggl(\sqrt{1-\frac{2kM}{{c_0}^2r}}
-1\biggr).
\eqno(102)
$$

The limit of expressions (101) and (102) for $c_0\to \infty$
leads to metric (96), which
corresponds to the perfectly rigid body model in the Newtonian sense. We call
such a body, wherein the longitudinal sound velocity is equal to the velocity
of light in vacuo, a relativistically rigid body [15]. Therewith expression
(101) exactly coincides with the component $\gamma_{11}$ of
Schwarzschild's metric
in the standard form, and the component $g_{00}$ of this metric
is obtained from
(102) if one expands $\exp(\nu)$ into  a series and retainis only the
first order
infinitesimal with respect to $(r_g/r)$.

Thus, for a sphericall symmetric NFR, whose basis is a relativistically rigid
body, and the acceleration corresponds to the Newtonian one, the metric has
the form (37), where $\nu$ is determined from (102) at the sound velocity
$c_0$
equal to that of light in vacuo, and $\lambda$, under the same conditions,
from (101). The calculation of the well-known GR effects in this metric only
slightly differs from the calculation using Schwarzschild's metric. A
difference emerges in the calculation of the pericentre displacement that is
5/6 of Schwarzschild's. The change of the direction of light ray, while
passing by the central body, coincides with Schwarzschild's. Hence, the
modified model corresponds to GR much closer than (96).

\begin{center}
{\bf 9. Conclusion}
\end{center}

Enumerate the basic results obtained in the paper.

1. An exact static solution has been found for the charge field in a
uniformly accelerated NFR realizable in Riemannian space-time. This solution
allows one, in principle, to seek a space-time structure and to find fields
from charged conductors of arbitrary shape. "Relativistic corrections" proved
to be small for positively charged bodies, and the expressions for the scalar
potential and the affine components of the field tensor coincide with the
conventional ones in Minkowski space. These corrections may be considerable
for negatively charged conductors or those being in an external electric
field.

{\sl However, the space-time metric both outside positively and
negatively charged
bodies is non-Euclidean}.

In present paper the cause of this phenomenon has been found and simplest
experiments allowing one to confirm or rule out the effects being predicted.

2. An exact expression has been found for the field and geometry of space-time
outside a charged metallic ball. If the Coulomb repulsion force between
charges on the ball surface is equal to the Newtonian attraction force and a
similar equality is fulfilled between test particles outside the ball, then
from our solution, to a high accuracy, follows Reissner-Nordstr{\"o}m's well-
known exact solution characterizing the electrovacuum static spherically
symmetric consistent solution to Einstein's and Maxwell's equations for
electrically charged point mass field.

If the charged ball radius vanishes, and the electron charge is "smeared"
over the sphere, then we shall obtain a point charge field.

{\sl In the present paper an exact expression has been found for
the energy of a
point charge $W$, which is equal to $W=2mc^2$ and independent of the quantum
density of charge. This expression eliminates the main difficulty of
classical and quantum electrodynamics leading to an infinite proper energy
of a point charge.

3. It is proved that the field energy ouside the cylinder of length $h$ and
infinite radius is a finite quantity, which differs from the classical value
of the charged filament energy that is infinitely large. In a close packing
of charges along the filament we have an expression for the energy
$$
W=\frac{\g hmc^2}{e}=Nmc^2,
$$
where $N$ is the number of electrons along the filament length.

{\sl Thus, for
three kinds of symmetry (flat, spherical and cylindrical) the field energy
created by point charges does not diverge, as in the classical case, but is
determined by the rest energy of charges creating the field. Therewith the
quantity of charge drops out of the formulae for the energy}.

4. A space-time metric which only slightly differs from Schwarzschild's has
been found on the basis of a sphericall symmetric NFR, wherein the sound
velocity is equal to that of light and the acceleration (in tetrads)
corresponds to the Newtonian one. The calculation of the well-known GR
effects from the found metric is close to the classical one. A difference
emerges only in the calculation of ther pericentre displacement that is 5/6
of Schwarzschild's.

As a rigid body (in the classical sense) was chosen to be a spherically
symmetric NFR and the Newtonian law of gravitation was considered to be
accurate,

{\sl then the space-time in such a model proved to be Riemannian
with a flat spatial cross-section}.

Thus, in the present paper a connection between the Newtonian and Einsteinian
theories has been established to be deeper than generally assumed.

5. An existence of the space-time curvature stimulated the search for a
connection with Einstein's equations. An exact solution to Einstein-Maxwell's
system of equations with a "cosmological constant" has been found, where the
electromagnetic field energy-momentum tensor is used as a source in
Einstein's equations. The solution found describes charged dust being in
equilibrium in "parallel" homogeneous electric and gravitational fields. It
has been proved that dust particles having a proton charge should have a mass
of the order of a mass of a stable elementary black holes - "maximons" [8].

6. The model being proposed has eliminated the main contradiction between the
charged particles being pointlike and their infinite proper energy. According
to the aforesaid, not only gravitational but even simplest electrostatic
fields curve space-time geometry.

This work is an attempt to establish a link between the space-time curvature
and the classical fields of bound structures.


\end{document}